\documentstyle[prb,aps,epsfig,floats]{revtex}
\input{epsf}

\begin{document}
\draft
\title{Measurements of critical current diffraction patterns \\
in annular Josephson junctions}

\author{Andreas Franz, Andreas Wallraff, and Alexey V. Ustinov}
\address{Physikalisches Institut III, Universit\"{a}t
  Erlangen-N\"{u}rnberg, D-91058 Erlangen, Germany}
\date{\today}
\wideabs{
\maketitle

\begin{abstract}
We report systematic measurements of the critical current versus
magnetic field patterns of annular Josephson junctions in a wide
magnetic field range.  A modulation of the envelope of the pattern,
which depends on the junction width, is observed.  The data are
compared with theory and good agreement is found.
\end{abstract}

}

Large area Josephson junctions are intriguing objects for performing
experiments on nonlinear electrodynamics.  In particular, the
propagation of solitons, also called Josephson vortices or fluxons,
has attracted a lot of attention and has been studied in
detail.\cite{ustinov98} Recently large area Josephson junctions have
been proposed to be used as efficient radiation and particle
detectors.\cite{esposito,nappi97a,cristiano99} In such junctions, the spatial
dependence of the phase difference between the superconducting
electrodes is an important characteristic that determines the junction
properties.

In an annular Josephson junction, magnetic flux quanta threading one
superconducting loop but not the other, are trapped and stored in the
junction due to the fluxoid quantization.\cite{davidson85} This
property of the system offers the unique possibility to study fluxon
dynamics in the absence of collisions with
boundaries.\cite{mclaughlin78} In the annular junctions
proposed\cite{nappi97a,cristiano99} for radiation and particle
detection, trapped vortices are useful to suppress the critical
current of the junction in order to allow for a stable bias point in
the subgap region of the junction current-voltage characteristic.

In this report, we present systematic measurements of the critical
current of annular Josephson junctions in dependence on the externally
applied in-plane magnetic field.  The critical current $I_c$ of a
junction without trapped fluxons is at maximum when no magnetic fields
are present.  In the presence of a magnetic field this maximum
superconducting current is reduced.  Magnetic fields can be due to the
bias current applied to the junction (self-fields), due to flux
trapped in the junction itself or its leads (Josephson or Abrikosov
vortices, respectively), or they can be applied externally.  The
modulation of the critical current with the external field is often
called a critical current diffraction pattern.  We investigate these
patterns for annular junctions of various dimensions in a wide range
of magnetic fields.

\begin{figure}[btp]
\epsfig{file=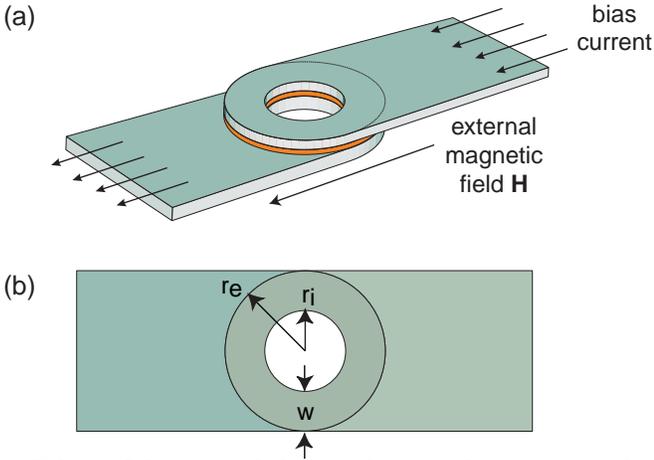, width=\columnwidth}
\caption{Schematic of the annular Josephson junction biased in the
Lyngby-geometry\cite{davidson85} (a) and its dimensions (b).}
\label{annular}
\end{figure}

We present experimental data on five annular Josephson junctions with
the same external radius $r_e = 50 \,{\rm \mu m}$ but different inner
radii $r_{i}$ ranging from 30 to 47 $\mu$m, see second and third
column of Tab.~\ref{tab:ringeNappi}.  Hence the width $w = r_{e} -
r_{i}$ of the junctions is ranging from $3$ to $20\, \rm{\mu m}$.  The
junction geometry is shown in Fig.~\ref{annular}.  All junctions have
been prepared on the same chip using Hypres technology\cite{Hypres}
with a nominal critical current density of $j_{c} = 100 \,
\rm{A/cm^{2}}$.  Accordingly, the Josephson length is approximately
$30 \,{\rm \mu m}$ at 4.2 K.

\begin{table}[btp]
    \caption{Geometrical parameters and fitted values 
    of the measured annular Josephson junctions.}
    \begin{tabular}[bthp]{c c c c c c c}
junction \# &
$r_i$  [$\mu$m] &
$\delta = r_i /r_e$ &
$\Delta H' / \Delta H $ &
$2 r_e / w$ &
$\tilde{\delta}$ &
$\Delta r$ $[\rm{\mu m}]$ 
\\ \hline 
A & 47& 0.94 & -    & -    & 0.96 & 0.5   \\ 
B & 45& 0.9  & 22.9 & 20.0 & 0.92 & 0.5   \\ 
C & 42& 0.84 & 13.2 & 12.5 & 0.88 & 1.0   \\
D & 35& 0.7  &  6.5 &  6.7 & 0.72 & 0.6   \\
E & 30& 0.6  &  4.8 &  5.0 & 0.62 & 0.5   \\
\hline \hline
junction \# &
$\tilde{H}_{0}$ [Oe] & 
$\tilde{H}_{0} / H_{0}$ &
$\tilde{\Lambda}$ $[\rm{nm}]$  & & \\
\hline
A & 0.319   & 0.65  & 208 & & &\\
B & 0.346   & 0.703 & 193 & & &\\ 
C & 0.321   & 0.646 & 210 & & &\\   
D & 0.405   & 0.821 & 165 & & &\\ 
E & 0.376   & 0.765 & 177 & & &\\ 
\end{tabular}
\label{tab:ringeNappi}
\end{table}

\begin{figure}[tbp]
\begin{center}
\epsfig{file=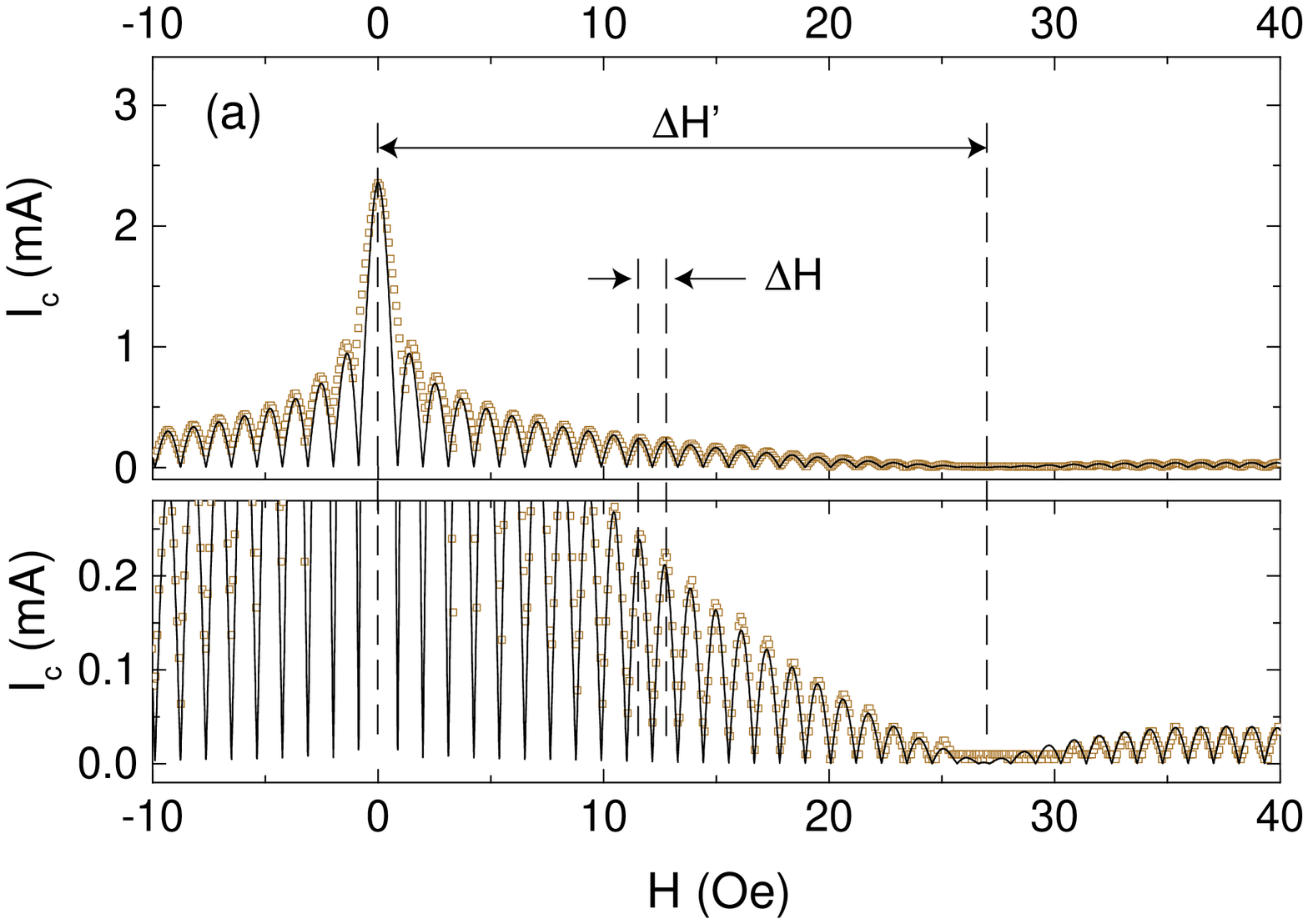, width=\columnwidth}
\epsfig{file=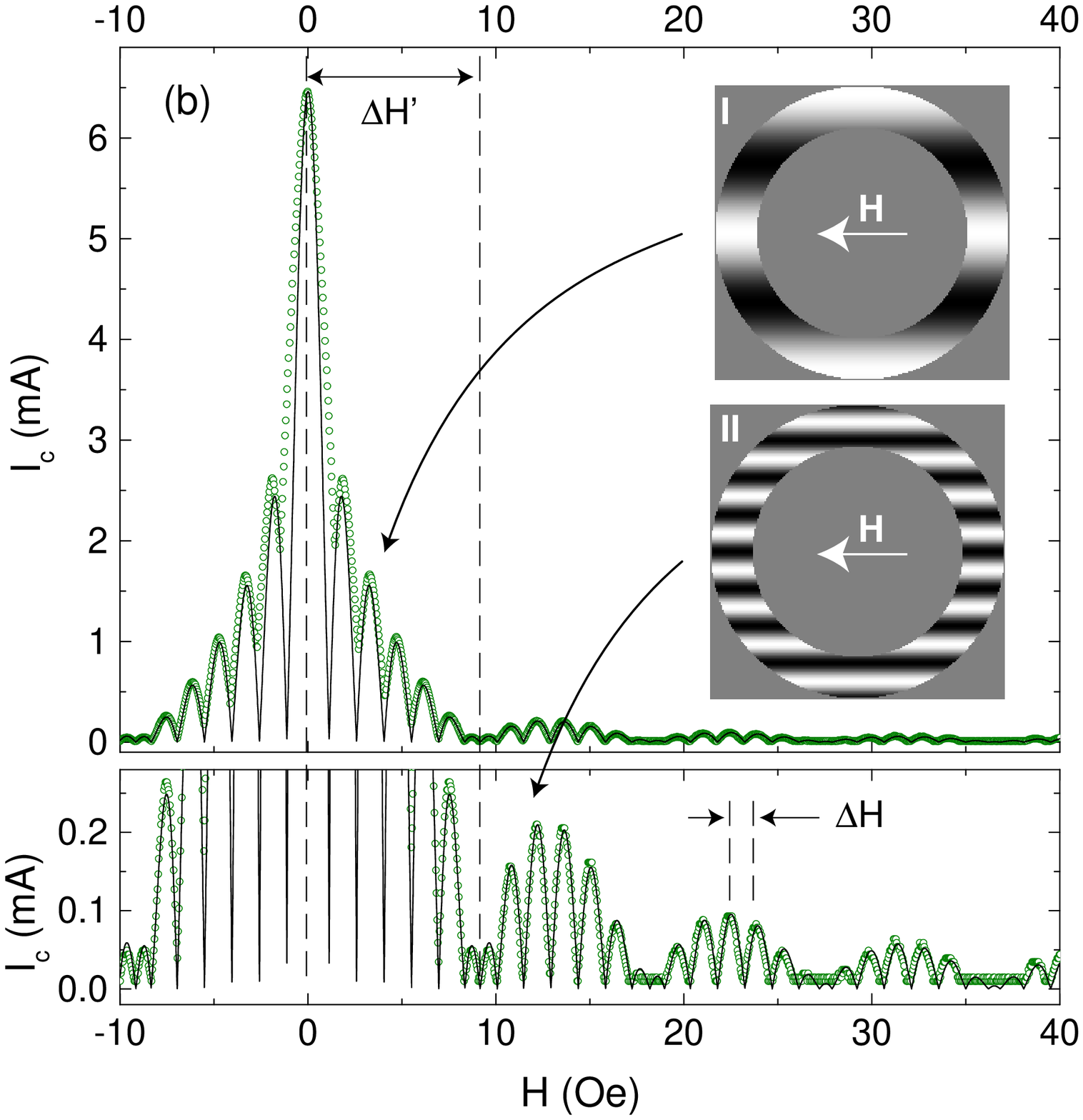, width=\columnwidth}
\caption{Critical current diffraction patterns of (a) junction B and
(b) junction D at $4.2\,{\rm K}$.  Dots are experimental data, the
solid line is theory according to Eq.~(\ref{eq:nappi}).  For better
visibility the low current region is also plotted on an enlarged
scale.  The two field modulation periods $\Delta H$ and $\Delta H'$
are indicated in each plot.  The insets $I$ and $II$ of plot b)
display the supercurrent distribution in junction D at the magnetic
fields indicated by arrows; light (dark) regions correspond to current
in positive (negative) direction.}
\label{1416noflux}
\end{center}
\end{figure}

In Fig.~\ref{1416noflux} the critical current diffraction patterns of
the two junctions B and D, being representative for the set of
measured samples, are shown.  Obviously, a strong dependence of the
pattern on the junction width is observed.  As expected, the critical
current of the junction at zero field scales with the junction size as
$I_{c} = j_{c} 2 \pi (r_{e}^2 - r_{i}^2)$.  Measuring the diffraction
patterns in a wide range of magnetic field, two characteristic
modulation scales of the critical current are observed.  The pattern
having a small magnetic field period $\Delta H$ has an envelope of the
larger period $\Delta H'$ which depends strongly on the width of the
junction (compare Figs.~\ref{1416noflux}a and b).

The observed critical current diffraction patterns can be
qualitatively understood in the following way.  The modulation of the
period $\Delta H$ is due to the penetration of magnetic flux in the
direction perpendicular to the external magnetic field.  This period
is inversely proportional to the diameter of the junction: $\Delta H
\propto 1/ (2 r_{e})$.  This is analogous to the standard case, where
$\Delta H$ is proportional to the reciprocal junctions length in the
direction perpendicular to the magnetic field.\cite{barone} The minima
of the modulation of the period $\Delta H'$ occur, when the magnetic
flux penetrates the junction strongly also along the \emph{width} of
the junction.  Therefore, the period $\Delta H'$ of the second
modulation is proportional to $1/w$.  By calculating the
ratio
\begin{equation} 
    \label{eq:over} 
    \frac{\Delta H'}{\Delta H} = \frac{2 r_{e}}{w}\;,  
\end{equation}
for the different junctions, this simple prediction can be
quantitatively compared with experiment.  As can be seen from the
forth and fifth column of Tab.~\ref{tab:ringeNappi}, Eq.\
(\ref{eq:over}) is quite accurately fulfilled for our
junctions.\cite{notejuncA}

The described effect can be illustrated by plotting the supercurrent
density $j_{s}$ at different magnetic magnetic fields versus the
junction coordinates.  At the magnetic field $H = 3.25 \, {\rm Oe} <
\Delta H'$, approximately two and a half flux quanta penetrate into
the junction cross section $2 r_{e}$, as shown in the inset I of
Fig.~\ref{1416noflux}b.  At the larger field $H = 12.2 \, {\rm Oe} >
\Delta H'$, more than one flux quantum penetrates the width cross
section of the junction (see inset II of Fig.~\ref{1416noflux}b). 
Thus, after each period $\Delta H'$, one additional flux quantum has
penetrated the width of the junction.  We note here, that the
spatial distribution of the supercurrent density could also be
measured in experiment.\cite{keil96}

Several approaches to calculate the critical current diffraction
patterns of annular Josephson junctions have been published
earlier.\cite{martucciello96a,martucciello96b,nappi97} Mainly, two
different cases have been considered, the long annular Josephson
junction with a circumference $2\pi \overline{r}$ larger than the
Josephson length $\lambda_J$ and the small annular Josephson junction
where $2\pi \overline{r} < \lambda_J$,\cite{martucciello96a} where 
$\overline{r} = (r_{i}+r_{e})/2$ is the mean radius of the junction. 

The most complete theoretical description of the
critical current diffraction pattern $I_{c}(H)$ of small annular
junctions of arbitrary width and number of trapped fluxons $n$ is
presented by Nappi in Ref.~\onlinecite{nappi97}. The dependence of the 
critical current $I_{c}$ on the magnetic field $H$ is given by the 
formula
\begin{equation}
    \label{eq:nappi} 
    I_c = I_0 \left|\frac{2}{1 - \delta^2}\int_{\delta}^1
    x J_n\left(x \frac{H}{H_0} \right) dx \right| \;, 
\end{equation} 
where $J_n$ is $n$-th Bessel function of integer order, $\delta = r_i
/ r_e$ is the ratio of inner radius $r_i$ to outer radius $r_e$ and
$I_{0}$ is the maximum superconducting current at zero field. The 
field
\begin{equation}
    H_0 = \Phi_0 / (2 \pi r_e \mu_0 \Lambda)
    \label{eq:Hnormparam}
\end{equation}
is the characteristic magnetic field; $\Phi_{0}$ is the flux quantum,
$\mu_{0}$ is the vacuum permeability and $\Lambda$ is the effective
magnetic thickness.\cite{weihnacht}

For $n=0$, the two extreme cases $\delta \rightarrow 1$ (see
Ref.~\onlinecite{martucciello96a}) and $\delta \rightarrow 0$
(Ref.~\onlinecite{barone}) of Eq.\ (\ref{eq:nappi}) have been
discussed in the literature.  The predictions of Eq.\ (\ref{eq:nappi})
have also been compared to experiments in a relatively small magnetic
field range.\cite{cristiano99,martucciello96b,nappi98}  To our
knowledge, there has been no systematic comparison of the theory with
experimental data for different junction width in a large field range. 
Our intention here is to perform such a comparison.

In Figure \ref{1416noflux}, our experimental data are fitted to
Eq.~(\ref{eq:nappi}).  In the fitting procedure the values of both
$H_0$ and $\delta$ are determined.  Subsequently, the quantities
acquired from the fits are labeled by a tilde ($\tilde{H}_{0}$,
$\tilde{\delta}$).  For the fit, the initial value of $\tilde{\delta}$
is calculated from the designed geometry of the junction; the initial
$\tilde{H_{0}}$ is calculated according to Eq.~(\ref{eq:Hnormparam})
assuming the reasonable value of $200 \, \rm{nm}$ for the magnetic
thickness $\Lambda$.  Then, the best fit is found by iteratively
adjusting $\tilde{H_{0}}$ and $\tilde{\delta}$.  The value of
$\tilde{H}_{0}$ predominantly determines the small period of the
critical current modulation $\Delta H$, whereas $\tilde{\delta}$
determines the large modulation scale $\Delta H'$.  This fact is in
agreement with the qualitative discussion above.  As can be seen from
Fig.~\ref{1416noflux}, excellent agreement between theory and
experiment is found.  The parameters $\tilde{\delta}$ and
$\tilde{H}_{0}$ determined from the best fits to the data of junctions
A to E are quoted in Tab.~\ref{tab:ringeNappi}.

Comparing the values of $\delta$ and $\tilde{\delta}$ in
Tab.~\ref{tab:ringeNappi}, we find that $\tilde{\delta} < \delta$ for
all junctions. This small but systematic deviation can be explained by
assuming a symmetric deviation $\Delta r$ of the junction radii
from their designed dimensions (e.g., due to the photolithographic 
procedure during the preparation). Using this assumption,
$\tilde{\delta}$ can be expressed as
\begin{equation}
  \label{eq:fitradii}
  \tilde{\delta} = \frac{r_i + \Delta r}{r_e - \Delta r}\;.
\end{equation}
From the fits we find that $\Delta r$ varies between $0.5$ and $1.0 \,
\rm{\mu m}$ (see Tab.~\ref{tab:ringeNappi}).  This size correction can
also be explained as due to a slight over-etching of the trilayers
during sample fabrication, that results in a small reduction of the
sample size.  The obtained $\Delta r$ values agree with
the size tolerance quoted by Hypres.\cite{Hypres}

According to theory, the quantity $H_{0}$ does only depend on the
outer junction radius $r_{e}$ and hence should be identical for all
junctions measured.  Instead, we find values of $\tilde{H}_{0}$ from
the fits that slightly vary from junction to junction, see
Tab.~\ref{tab:ringeNappi}.  Using Eq.~(\ref{eq:Hnormparam}) the
magnetic thickness $\tilde{\Lambda}$ can be calculated from
$\tilde{H}_{0}$.  The values of $\tilde{\Lambda}$ obtained for each
junction are quoted in the last column of Tab.~\ref{tab:ringeNappi}. 
The average magnetic thickness is $\tilde{\Lambda} = 191 \pm 18 \,
\rm{nm}$, which is in good agreement with the value of $\Lambda
\approx 2 \lambda_{L}$ in the thick film limit, yielding a London
penetration depth of $\lambda_{L} \approx 95 \, \rm{nm}$.

The scatter observed in $\tilde{H}_{0}$ (or, equivalently, in
$\tilde{\Lambda}$) may be due to a small number $m$ of flux quanta
threading the holes of both junction electrodes simultaneously.  The
critical current diffraction patterns for different values $m$ are
very similar in their qualitative features, but may differ
quantitatively.  Preliminary experimental results show that, upon
cooling the junction from the normal to the superconducting state in a
small residual magnetic field a large number of times and measuring
the resulting critical current versus magnetic field, slightly
different diffraction patterns depending on the value of $m$ are
observed.  In such measurements we have only observed three different
diffraction patterns, despite repeating the described procedure a
large number of times.  This strongly suggests that this effect is 
due to magnetic flux threading the junction loop perpendicular to the 
substrate.

At small fields, we observe a systematic deviation of the calculated
patterns from the experimental ones.  In particular, the first minimum
of the critical current appears at larger field values than predicted
by the theory.  Moreover, the critical current at the first minimum
does not fall to zero.  Both facts are to be expected for junctions
that are not really small in comparison with $\lambda_{J}$.  Indeed,
the dimensions of our junctions are slightly larger than
$\lambda_{J}$.  This leads to a inhomogeneous penetration of the
magnetic field into the junction at low fields, resulting in an
increase of the field value $H$ at which the the first minimum of the
pattern is observed.  The analogous effect is observed in conventional
long Josephson junctions.\cite{barone,pagano}

\begin{figure}[tbp]
  \begin{center}
    \epsfig{file=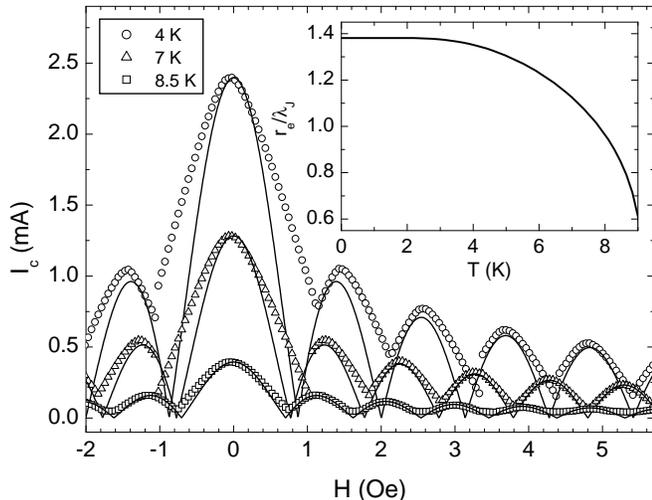, width = \columnwidth}
    \caption{Critical current diffraction patterns of junction B at
    temperatures between $4 K$ and $8.5 K$.  Dots are experimental data,
    solid line is theory. In the inset the calculated normalized external 
    junction radius $r_{e}/\lambda_{J}$ is plotted versus temperature.}
    \label{fig:temp}
  \end{center}
\end{figure}

At higher temperatures, the Josephson length $\lambda_{J}$
increases\cite{barone} and, hence, the effective size of the junction
decreases.  In the inset of Fig.~\ref{fig:temp}, the calculated
normalized external junction radius $r_{e}/\lambda_{J}$ is plotted
versus temperature, taking into account the temperature dependence of
both the critical current density $j_{c}(T)$ and the London
penetration depth $\lambda_{L}(T)$.\cite{barone} At $T > 7.8 \,
\rm{K}$ the normalized radius drops below unity.  Therefore, at higher
temperatures a better agreement between experimental data and theory
can be expected at low fields.  This is illustrated in
Fig.~\ref{fig:temp}, where the experimental critical current
diffraction pattern of junction B is plotted together with a fit for
the temperatures $T = 4.0, \, 7.0, \, 8.5 \, \rm{K}$.  The fit is made
keeping $\tilde{\delta}$ constant for all $T$ and adjusting
$\tilde{H}_{0}$.  At elevated temperatures, both the position of the
first minimum and the modulation depth of the critical current at
small fields show better agreement with the theoretical prediction.

\begin{figure}[tbp]
  \begin{center}
    \epsfig{file=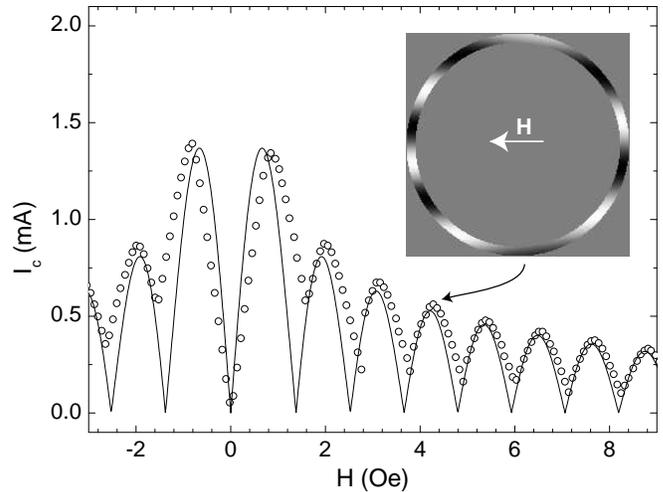, width= \columnwidth}
    \caption{Diffraction pattern of junction B with one trapped 
    fluxon at $4.2 \, \rm{K}$. Dots are experimental data,
    solid line is theory. The inset displays the supercurrent 
    distribution in the junction at the magnetic field indicated by 
    the arrow.}
    \label{fig:14oneflux}
  \end{center}
\end{figure}

We have also measured junctions with a single fluxon trapped in the
junction barrier ($n=1$).  As an example, the critical current
diffraction pattern of junction B for $n=1$ at 4.2 K is shown in
Fig.~\ref{fig:14oneflux}.  Taking the same fitting parameters as for
the case of no trapped fluxon ($n=0$), we find as good agreement
between the theory and the experimental data as before.  The slight
differences between the fit and the experimental data are, again, due
to the dimensions ($r_{e}>\lambda_{J}$) of the junction.
In the inset of Fig.~\ref{fig:14oneflux} the supercurrent distribution
in junction B at $H = 1.71 \, \rm{Oe}$ calculated according to
Eq.~(\ref{eq:nappi}) is shown.  Obviously at this field a number of
vortex anti-vortex pairs have penetrated into the junction but the
width of the junction is not fully penetrated (compare
Fig.~\ref{1416noflux}b, inset II).  The symmetry of the current
distribution in the junction is broken due to the presence of the
trapped vortex.  Similar current distributions in the presence of
trapped vortices have also been observed in experiment.\cite{keil96}

It is worth to point out that good agreement between theory and
experiment in the large field range is found for junctions of a
diameter substantially larger than $\lambda_{J}$.  At low fields the
theory\cite{nappi97} describes well the experiments with $r_{e} <
\lambda_{J}$, as confirmed by our measurements at higher temperatures. 
Thus, the magnetic properties of the junction are determined rather by
the junction radius than by the junction circumference, as already
pointed out in Ref.~\onlinecite{martucciello98}.

In summary, we have systematically measured the critical current
diffraction patterns of a number of annular junctions of different
width, with and without trapped fluxons, in a wide magnetic field
range and at different temperatures.  The experimental data show a
pronounced width dependence that is explained accurately using the
existing theory.  In particular, a modulation of the envelope of the
critical current diffraction pattern is observed for junctions of
large width.  The period of this modulation depends very
sensitively on the normalized junction size described by the parameter
$\delta$.  The method of our data analysis is accurate enough to
detect a small reduction of the size of the junction due to the
fabrication process.


\begin{thebibliography}{10}
\bibitem{notejuncA}For junction A, $\Delta H'$ was not evaluated.  Due
to its small width, the field necessary to observe the first minimum
of the envelope of the diffraction pattern was above the range of our
measurements.
\end{thebibliography}
\end{document}